\documentclass[12pt,preprint]{cmlapss}
\newcommand{\beq}{\begin{equation}}
\newcommand{\eeq}{\end{equation}}
\newcommand{\beqn}{\begin{eqnarray}}
\newcommand{\eeqn}{\end{eqnarray}}

\begin{document}
       
%
\title{\bf The Effect of Impact Parameters on the
Formation of Massive Black Hole Binaries in Galactic Mergers}

\author{Yu-Heng Ho$^{1}$, Ing-Guey Jiang$^{1,2}$,Yu-Ting Wu$^{3}$}

\affil{
{$^{1}$Institute of Astronomy,}
{National Tsing-Hua University, Hsin-Chu, Taiwan}\\
{$^{2}$Department of Physics,} 
{National Tsing-Hua University, Hsin-Chu, Taiwan}\\
{$^{3}$National Astronomical Observatory of Japan, 
Mitaka, Tokyo 181-8588, Japan}\\
}
\email{jiang@phys.nthu.edu.tw}
\begin{abstract}
By employing N-body simulations,
we inestigate the formation of massive black hole binaries (MBHBs)
through the sinking of two massive black holes (MBHs)
during galaxy mergers. 
With different impact parameters and different central stellar density 
of the progenitor galaxies, we analyze the orbits of the MBHs 
from the beginning of the merger 
until the time when the bound MBHB forms.
Contrary to the previous theory that the timing of the dual MBHs entering
their dynamical radius is similar as the timing of the formation 
of the bound MBHB, 
we find that these two timings could deviate when the 
central stellar density of the progenitors galaxies are lower.
On the other hand, when the 
central stellar density of the progenitor galaxies is
higher and the mergers have small impact parameters, 
each MBHs would move directly into the 
core radius of the other progenitor galaxies, 
and therefore cause a variation in the timings of the MBHB formation.
\end{abstract}

\section{Introduction}\label{sec:intro}

Massive black holes (MBHs), 
viewed as the central engines of active galactic nuclei (AGNs), 
are now generally considered a fundamental component of massive galaxies 
due to their ubiquitous detection in these galaxies.
It is known that many AGNs are found 
at very high redshift.
For example, 
a luminous quasar ULAS J112001.48+064124.3 was reported 
to host a MBH of $2 \times 10^9 M_\odot$ at redshift $z \sim 7.085$ 
\citep{2011Natur.474..616M}, 
and a quasar ULAS J134208.10+092838.61 was found to host a MBH with a mass
of $8 \times 10^8 M_\odot$ at redshift $z \sim 7.54$ 
\citep{2018Natur.553..473B}.
These results imply that MBHs shall exist in the very early Universe. 

In the hierarchical formation scenario of galaxies in the Local Group, 
such as the Milky Way and the Andromeda Galaxy, 
are expected to have accreted their mass through mergers 
of progenitor galaxies 
\citep{2000MNRAS.319..168C, 2013ApJ...770...57B, 2015ARA&A..53...51S, 2019MNRAS.488.3143B}. 
Massive elliptical galaxies are also speculated to form through major mergers 
of galaxies\citep{1996IAUS..171..181B, 2006MNRAS.366..499D, 2007A&A...476.1179B, 2017MNRAS.467.3934D}.
Because MBHs are ubiquitously found in these massive galaxies, 
which very likely have experienced mergers in the past,
it is important to study the evolution of pair MBHs during merger, 
particularly the formation of massive black hole binaries (MBHBs).

The formation of MBHBs is also important in terms of the connection 
between the AGN periods and galaxy mergers.
The first thorough discussion by \citet{1980Natur.287..307B} 
showed that the motion of black hole binaries in a newly merged galaxy 
may cause the bending or precession features in some AGN jets. 
However, due to observational limitations, such as the limitation of angular resolutions, 
the number of resolved MBHBs is still small\citep{2004ApJ...602..123M, 2011MNRAS.410.2113B, 2016MNRAS.459..820T} 
and thus the evolution of MBHBs in the Universe still remains largely unclear. 

The formation of MBHB was divided into four stages 
by \citet{2002MNRAS.331..935Y}, including 
the dynamical friction stage, the non-hard binary stage, 
the hard binary stage, and the gravitational radiation stage. 
In the dynamical friction stage, 
two MBHs that are originally located at the center of the progenitor galaxies 
lose energy due to dynamical friction and therefore migrate towards 
the center of the merged system\citep{1996NewA....1...35Q, 2002MNRAS.331..935Y, 2003AIPC..686..201M, 
2016MNRAS.461.1023B}. 
These two MBHs fall from a kpc-scale separation down to a pc-scale separation 
and eventually are bound by gravity. 
This marks the formation of the MBHB 
and the end of the dynamical friction stage. 
The timescale of this stage strongly depends 
on the geometry of the MBHs' orbits and 
the stellar density surrounding the MBHs. 
In addition, this is also the stage 
at which dual or offset AGNs can 
appear\citep{2010ApJ...715L..30L, 2015ApJ...806..219C, 2019ApJ...882...41H}. 
Therefore, estimating the timescale of the dynamical friction stage is crucial in terms of predicting the lifetime of dual AGNs.

The non-hard binary stage is the stage where the MBHB keeps losing energy until 
its orbital speed becomes comparable to the velocity dispersion of the stars in the  
central galactic region\citep{1980Natur.287..307B}. 
In this stage, the MBHB loses its orbital energy through interactions with nearby stars, 
and therefore its semi-major axis decreases. This process is so-called 
\textit{dynamical hardening}. 
During this process, 
dynamical friction gradually becomes less dominant because 
the velocity of the MBHs increase.

The third stage is the hard binary stage at which the orbital speed of the MBHB is comparable to
the velocity dispersion of the central stars\citep{1980Natur.287..307B}, and thus the MBHB is considered a hard binary. 
In this stage, dynamical friction no longer plays an important role since only a small portion 
of stars can get sufficiently close to the hard MBHB. 
Moreover, the MBHB would clear off a so-called "loss cone" region of stars in the phase space
\citep{1996NewA....1...35Q, 2004ApJ...606..788M, 2006RPPh...69.2513M}. 
The timescale of this stage depends on the re-population of stars into the loss cone.

Finally, in the gravitational radiation stage, when enough stars can diffuse into the loss cone, 
the MBHB continues to harden to the point where gravitational wave radiation takes over 
as the main mechanism subtracting the orbital energy. 
Eventually, the MBHB will coalesce\citep{1996NewA....1...35Q}.

Numerical simulations are powerful tools to investigate the formation of MBHBs
\citep{1996NewA....1...35Q, 2006ApJ...642L..21B, 2009ApJ...695..455B, 2012ApJ...749..147K}. 
However, due to the wide range in spatial resolution 
and the involvement of various physical mechanisms, 
it is very difficult to trace the evolution of 
the MBHs from the very beginning, 
when the MBHs were originally located at the centers of each progenitor galaxy, 
down to the scales where gravitational wave 
radiation drives the MBHB to 
coalescence, and leads to the formation of 
a more massive MBH\citep{2016ApJ...828...73K}. 
As the total number of particles $N$ is one of the important parameters in numerical simulations, 
\citet{1997ApJ...478...58M} performed N-body simulations 
with $2\times10^3 \le N \le 2.56\times10^5$ to study the effects of $N$ 
on the formation and evolution of MBHBs during galaxy mergers. 
They found that the evolution timescale of an MBHB is independent of $N$ 
until the separation of the MBHB reaches a critical value. 
\citet{2006ApJ...642L..21B} further used 
$2.5\times10^4 \le N \le 10^6$ to study the evolution of MBHBs 
and found that the hardening rate 
(the rate at which the semimajor axis of MBHB decreases) 
is independent of $N$. Since hard binaries are formed at parsec 
separations\citep{1996NewA....1...35Q, 2002MNRAS.331..935Y}, 
the above results suggest that N-body simulations with $N \sim $ 
tens-of-thousand particles should be sufficient to study the orbital evolution of the MBHBs from kpc to pc scale.

As shown in \citet{2002MNRAS.331..935Y},
whether an MBHB could form within a Hubble time 
would depend on the mass ratio of two MBHs. 
Using N-body simulations, \citet{2012ApJ...749..147K} found that 
the timescale for forming a MBHB is well within a Hubble time 
when it forms by the merger of two equal-mass MBHs. 
However, for a pair of MBHs with a small mass ratio,
the timescale is too large to form a MBHB within a Hubble time. 
In addition, it was demonstrated that the timescale of the 
dynamical friction stage 
is associated with the stellar density profile around MBHs. 
Using the analytical method and N-body integrations, 
\citet{2017ApJ...840...31D} found that the dynamical friction timescale 
can be long 
for shallower central stellar density cusps($\rho < r^{-1}$), 
and minor mergers with mass ratio $q < 10^{-3}$ 
could result in the less-massive MBH stalling at a distance 
of one-tenth the influence radius of the more-massive MBH.

To understand the effects of gas on the dynamical evolution of MBHBs, 
\citet{2017MNRAS.471.3646P} used hydrodynamical simulations 
to evaluate the process of the MBHBs formation. 
They found that under a smooth stellar potential of the galactic nucleus,
dynamical friction is the main mechanism driving the MBHs 
from kpc to pc scale
and gas does not play an important role 
when the initial gas fraction is 30\%. 
A higher gas fraction up to 50\% was considered by \citet{2017MNRAS.464.2952T}. 
They studied the evolution of an MBH pair 
(with a mass ratio of 0.2)
in a clumpy gaseous disk, and found that the orbital decay of the 
less-massive MBH 
could be either accelerated or delayed depending on whether a positive 
or negative torque was exerted on the less-massive MBH by 
the surrounding massive clumps. 
In some cases, he time-delay of the orbital decay could result in the formation of MBHBs taking
longer than 1 Gyr. 
Furthermore, on smaller scales($\leq$ 100 pc), 
\citet{2013ApJ...777L..14F} studied the evolution of two unequal mass MBHs 
in a clumpy gaseous disk, 
and showed stochastic behavior of the MBH orbits in their simulations. 
The less-massive MBH was perturbed by massive clumps and then scattered 
out of the disk plane. 
Its orbital decay timescales is therefore longer compared to 
the case where MBHs 
are in a smooth gaseous disk, and ranges from $\sim$1 to $\sim$50 Myr. 

In addition to examining the effects of gas, 
several studies have also been performed 
to investigate the effects of sub-grid physics, such as gas cooling, star formation, supernova feedback and feedback from black holes. 
For example, \citet{2015MNRAS.453.3437L} considered mergers of 
two disk galaxies with a gas fraction of $\sim$30\%, 
and found that the MBH orbits are not significantly affected 
by the recipes used to model the supernova feedback, 
but mainly affected by the gas clumpiness. 
In their simulations, the orbit decay timescale is $\sim$10 to 20 Myr. 
In addition, \citet{2017ApJ...838...13S} 
investigated the effects of feedback from black holes (BHs).
The timescale of the orbital decay of two BHs could be increased 
from $\sim$10 Myr 
to more than $\sim$300 Myr in their simulations when the feedback from black holes is considered. 
They suggested that the delay of the orbital decay is due to the 
{\it wake evacuation} effect, 
in which a hot bubble is created behind the less-massive BH and expels gas from its corresponding region. 
Therefore, the effect of the dynamical friction is reduced and the orbital decay is delayed.

In this work, we employed gravitational N-body simulations to study 
the formation of MBHBs over a range of different impact parameters and 
investigated the orbital evolution of the MBHB from 300 kpc 
to sub-kpc separations. 
In addition, we also built our progenitor galaxies with different core radii, 
so that we can study the effects of different impact parameters 
as well as core radii 
on the decay timescale of the MBHs orbit and the formation time of the MBHBs.

For simplicity, we only considered spherical progenitors
without rotation and did not consider the hydrodynamical effects, 
i.e. we focused on a major merger between two gas-poor spherical galaxies, 
although it is clear that gas may play a role in the decay timescale of the MBHs orbit 
based on the discussions in the previous paragraphs. 
If gas were included, clumps could form during the merging process 
and exert dynamical frictions onto the MBHs. 
Thus, the orbital decay of MBHs would be enhanced\citep{2017MNRAS.471.3646P, 2020ApJ...896..113L}. 
A qualitative study exploring the effects of gas in our simulations will be investigated in future work.
This paper is organized as follows. 
The model is described in Section~\ref{sec:simulation}. 
The results are presented in Section~\ref{sec:results}. 
The discussions about physical implications are mentioned in 
Section~\ref{sec:discussions}, 
and the study is concluded in Section~\ref{sec:conclusion}. 

\section{Numerical Simulations}\label{sec:simulation}

According to \citet{2001ApJ...563...34M}, the simulations of
galactic mergers with MBHs can be done by tree codes. 
As shown in their Fig.\,1, the time evolution of the stellar density 
and the stellar velocity dispersion near MBHs obtained by tree codes 
are completely consistent with those in direct N-body calculations. 
Only after the MBHB becomes a hard binary, 
the orbital evolution obtained from tree codes would start to differ 
from the results of direct N-body calculations. 

In this work, since we only focus on the orbital evolution 
of the MBHs before a hard binary is formed,
the publicly available tree-PM code GADGET-2 \citep{2005MNRAS.364.1105S} 
is adequate to perform simulations to achieve our goals. 
The initial galaxy models and the setup of mergers are presented below.

\subsection{Progenitor Models}\label{subsec:progenitor}

Our progenitor galaxy consists of three components, including a MBH, a stellar component, and a dark halo. 
The MBH is represented by one massive particle at the center of the galaxy.
The density profiles of the stellar component and the dark halo in
the galaxy follow spherical symmetric models 
in \citet{1990ApJ...356..359H} and \citet{1993ApJS...86..389H}. 
The density profile of the stellar component is 
\begin{equation}
    \rho_s(r) = \frac{M_s}{2\pi} \frac{r_s}{r} \frac{1}{(r + r_s)^3},
\end{equation}
where $M_s$ is the total stellar mass, and $r_s$ is
the characteristic radius, which determines the half-mass radius 
$r_{1/2} = (1 + \sqrt{2}) r_s$.
The density profile of the dark halo follows the form:
\begin{equation}
\rho_h(r) = \frac{M_h}{2\pi^{3/2}} \frac{\alpha}{r_0} \frac{\exp(-r^2/r_h^2)}{r^2 + r_h^2},
\end{equation}
where $M_h$ is the halo mass, $r_0$ is the cutoff radius, and $r_h$ is the characteristic radius of the profile.  
The normalization constant $\alpha$ is defined as
\begin{equation}
\alpha = \left\{ 1 - \sqrt{\pi} q \exp(q^2) \left[ 1 - \rm{erf}(q) \right]\right\}^{-1},
\end{equation}
where $q = r_h/r_0$. 
The initial velocities of particles were set to be in
dynamical equilibrium, which can be achieved by calculating the 
distribution function over the spherical shells 
as described in \citet{1982MNRAS.200..951B}.

Furthermore, in our galaxy model, the mass ratio between each component follows the statistical results in 
\citet{2013ARA&A..51..511K}, \citet{ 2010ApJ...710..903M} and \citet{2016MNRAS.462.1864L}.
The mass ratio between the MBH and the stellar part
was obtained through  
the scaling relation derived by \citet{2013ARA&A..51..511K}: 
\begin{equation}
\frac{M_{\bullet}}{10^9 M_{\odot}} = \left( 0.49^{+0.06}_{-0.05} \right) 
\left( \frac{M_s}{10^{11} M_{\odot}} \right)^{1.16 \pm 0.08}, 
\end{equation}
where $M_{\bullet}$ is the MBH mass, and $M_s$ is the total stellar mass 
of the galaxy. The relation between the stellar and dark halo mass is
\begin{equation}
\frac{M_s}{M_h} = 0.0564 \left\{ \left[ \frac{M_h}{7.66 \times 10^{11} M_{\odot}} \right]^{-1.06} + \left[ \frac{M_h}{7.66 \times 10^{11} M_{\odot}} \right]^{0.556} \right\}^{-1},
\end{equation}
where $M_h$ is the dark halo mass \citep{2010ApJ...710..903M, 2016MNRAS.462.1864L}.

Given that the MBH mass of $10^8 M_\odot$ is adopted in our galaxy model,
the total stellar mass of the galaxy $M_s$ is $2.5410 \times 10^{10} M_{\odot}$ and 
the dark halo mass $M_h$ is $8.7620 \times 10^{11} M_{\odot}$.
In addition, we adopt  $r_h = 10$ kpc and $r_0 = 25$ kpc in all our galaxy models. 
However, the $r_s$ is set to either 5 kpc or 8 kpc in two different galaxy models. 
With this setup, the stellar part is embedded within the dark halo. 
The number of particles representing the stellar component 
and dark halo in one galaxy are 127,050 and 730,164, respectively.

\subsection{Merger Setup}\label{subsec:setup}

\begin{figure}[tb]
\includegraphics[width=\columnwidth]{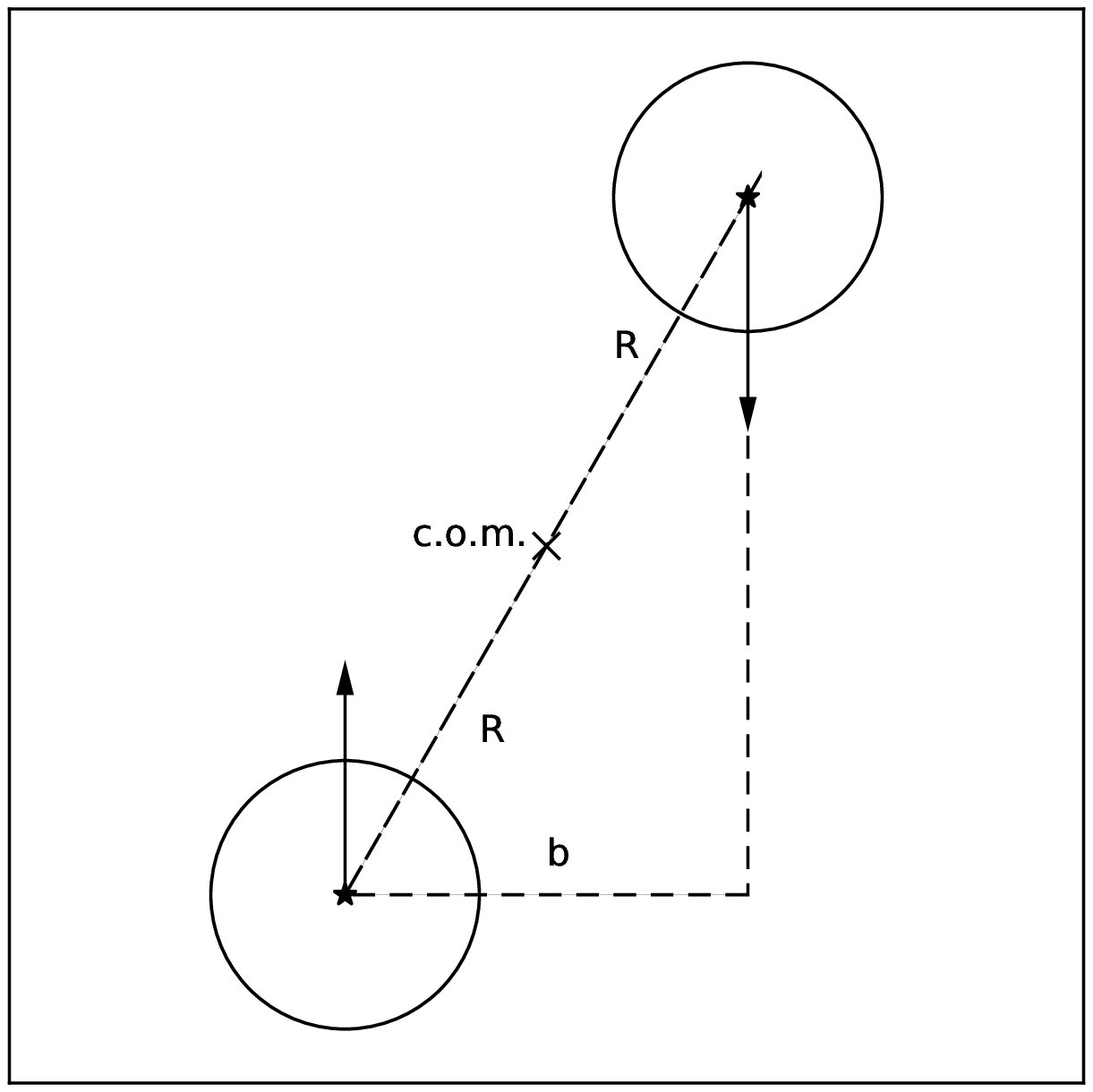}
\caption{An illustration of the merger simulation setup,
	 where c.o.m. is the center of mass, 
        $R$ is the orbital radius. 
	The galaxies are assumed to have parabolic encounters, 
	and the initial velocities are set to be anti-parallel. 
	The impact parameter $b$ is set to be  
        0, 10, 20, 30, 40, 50, 60, 70, and 80 kpc. 
	}
\label{scheme}
\end{figure}

Fig.~\ref{scheme} shows a schematic diagram of the initial setup of 
our simulations. 
The center of mass of two MBHs is placed at the 
coordinate origin in all merger simulations. 
The radial distance from one MBH to this center is called
orbital radius and denoted by $R$. It is set to be 150 kpc 
initially for both MBHs, 
so the initial separation between two MBHs is $2R=$ 300 kpc. 
We assume the approach of the progenitor galaxies to be 
a parabolic encounter. 
Therefore, the initial velocity of each progenitor would be 
\begin{equation}
    v_0 = \sqrt{\frac{GM}{2R}}, 
\end{equation} 
where $M$ is the total mass of each progenitor.
To investigate the effects of the impact parameter $b$, for each galaxy model, 
we perform nine simulations with impact parameters ranging from 0 to 80 kpc with a step of 10 kpc.

The gravitational softening lengths are set to be 
$10^{-5}$ kpc for the MBH particles,
$0.03$ kpc for the stellar particles, 
and $0.27$ kpc for the dark halo particles. 
The rest-frame is set to be at the center of mass of the two MBHs, 
so that the pair of MBHs would meet near the coordinate origin. 
We run the simulations until the MBHs form a bound binary, 
or until the evolution reaches 13 Gyr, i.e. about one Hubble time.

\section{Results}\label{sec:results}

In this section, 
we present the evolution of MBH sinking process, 
the MBHB orbits, and the stellar structure of the merged core in two different models.
Hereafter, we use "Model A" referring the mergers 
of the progenitor galaxies for which the characteristic radius $r_s$ is 5 kpc, 
and "Model B" for the mergers 
of the galaxies for which $r_s$ is 8 kpc. 
In addition, each model contains nine simulations with 
different impact parameters 
ranging from 0 to 80 kpc, as mentioned in Section~\ref{subsec:setup}. 
Then, we compared the results between Model A and Model B 
to see the effects of impact parameter and $r_\mathrm{s}$ of the progenitors 
on the formation and evolution of MBHBs.

\subsection{The MBH Sinking} \label{subsec:stages}

In addition to the orbital radius mentioned in Section~\ref{sec:simulation}, 
to better understand the orbital evolution of the two MBHs, 
we define two characteristic radii below, including 
the core radius $r_\mathrm{c}$ and the dynamical radius $r_\mathrm{dyn}$.
The evolution of each radius and the timings of the
MBHs reaching $r_\mathrm{c}$ and $r_\mathrm{dyn}$ will be presented 
in this section.
Note that the orbital radius of the two MBHs are identical 
in our simulations because we 
perform axisymmetric mergers with two equal-mass galaxies. 
Therefore, in the following we only present 
one orbital radius $R$ for both MBHs.

$r_\mathrm{c}$ and $r_\mathrm{dyn}$ are defined in the same reference frame 
as the orbital radius mentioned in Section~\ref{sec:simulation}, that is the spherical coordinate system 
centered at the center of mass of the two MBHs. 
The core radius $r_\mathrm{c}$ is useful to
trace the evolution of the stellar component during the merger, 
and is defined as
\begin{equation}
    r_\mathrm{c} = \frac{r_\mathrm{e}}{8}, 
    \label{r_c_def}
\end{equation}
where $r_\mathrm{e}$ is the effective radius, which is 
the radius that encloses half of the total stellar mass of two galaxies. 
This definition is the same as in \citet{1995MNRAS.276.1341J} and  \citet{2000ApJ...539L...9F} 
for the purpose of the measurements of velocity dispersion. 
On the other hand, we define the dynamical radius $r_\mathrm{dyn}$ 
as the radius that contains the stellar mass 
equaling to the MBHB mass $M_\mathrm{BB}$
\citep{1987gady.book.....B}:
\begin{equation}
    M_\star(r_\mathrm{dyn}) \equiv M_\mathrm{BB} = 2 M_\bullet.
    \label{r_infl_def}
\end{equation}

As the MBHs migrate inward during the merger process, 
the orbital radius $R$ could 
become comparable to $r_\mathrm{c}$ and $r_\mathrm{dyn}$.
However, since the MBHs are not on circular orbits,
their orbital radius $R$ may oscillate, which means that 
$R$ becomes smaller or larger than $r_\mathrm{c}$ and $r_\mathrm{dyn}$. 
This process can be repeated several times.
The final timing of the orbital radius $R$ being the same as 
the core radius $r_\mathrm{c}$ is defined 
as $t_\mathrm{c}$, i.e. 
\begin{equation}
    R(t_\mathrm{c}) = r_\mathrm{c}. 
    \label{t_c_def}
\end{equation}
Similarly, $t_\mathrm{dyn}$ is defined as the final timing when $R$ 
is the same as $r_\mathrm{dyn}$, i.e.
\begin{equation}
    R(t_\mathrm{dyn}) = r_\mathrm{dyn}. 
    \label{t_inf_def}
\end{equation}

\begin{figure}[tb]
\includegraphics[width=\columnwidth]{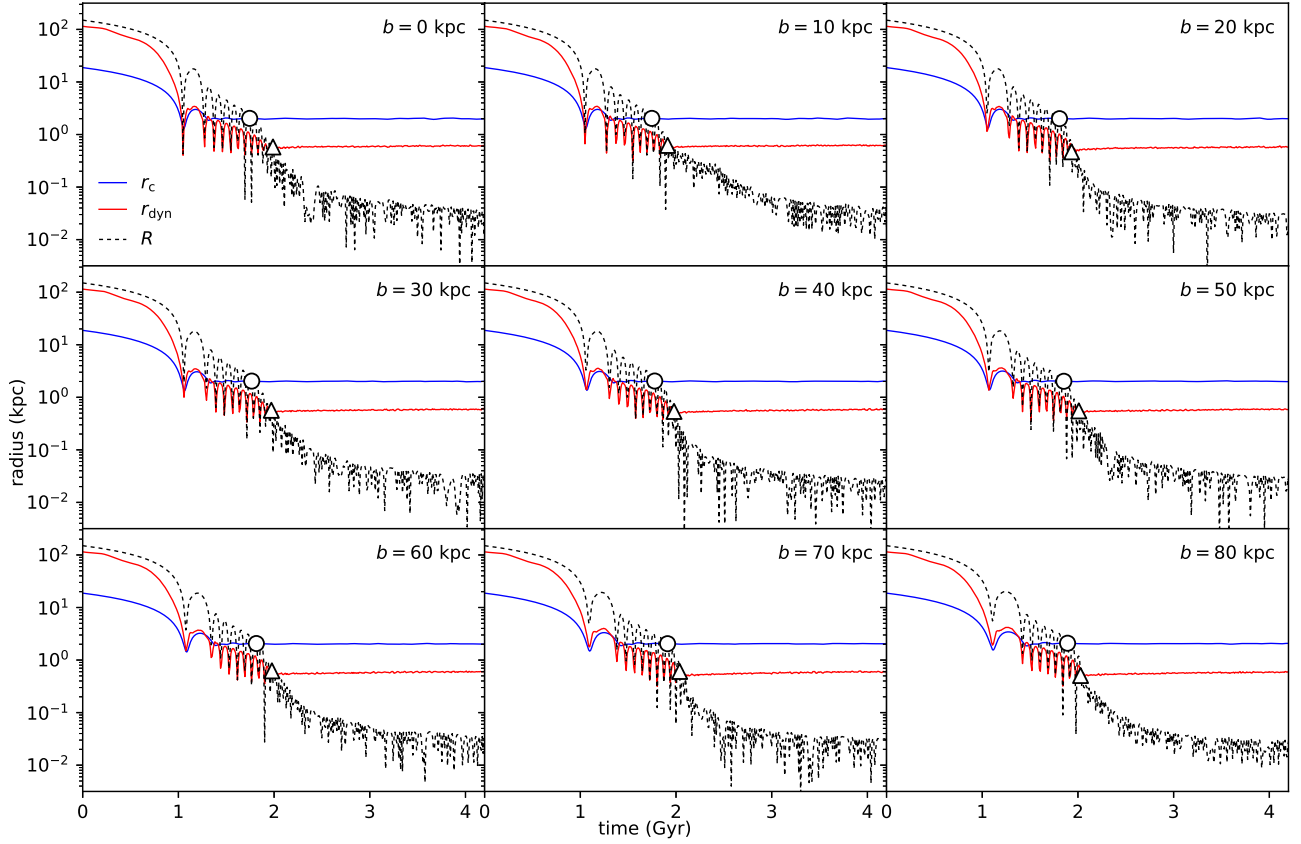}
\caption{Time evolution of core radius $r_\mathrm{c}$ (blue line),
dynamical radius $r_\mathrm{dyn}$ (red line), 
and orbital radius $R$ (black dotted line) for Model A.
The impact parameter $b$ is shown in 
the upper-right corner of each panel.
The circle and the triangle indicate 
$t_\mathrm{c}$ and $t_\mathrm{dyn}$, respectively. 
}
\label{R_c5}
\end{figure}

Fig.~\ref{R_c5} presents
the time evolution of the core radius $r_\mathrm{c}$ (blue line), 
the dynamical radius $r_\mathrm{dyn}$ (red line), 
and the orbital radius $R$ (black line) of Model A. 
In Fig.~\ref{R_c5}, nine panels beginning from top-left to 
bottom-right are for different impact parameters $b$. 
In general, the core radius $r_\mathrm{c}$ decreases 
from an initial value of $\sim$19 kpc, and settles to $\sim$ 2 kpc.
The dynamical radius $r_\mathrm{dyn}$ begins from around 100 kpc, and
settles to $\sim$ 1 kpc. 
Since the definitions of these radii are based on the cumulative mass 
profile of the merger remnant, the point where both radius 
become stationary shows that 
the merger remnant has reached a stabilized mass profile.
The orbital radius $R$ decreases from an initial value of 150 kpc,
and exhibits several damping oscillations. 
Finally it reaches saturation at $\sim 10^{-2}$ kpc.
As the orbital radius decreases, 
$t_\mathrm{c}$ and $t_\mathrm{dyn}$ can then be determined,
as indicated by the circle 
and the triangle in Fig.~\ref{R_c5}, respectively.
It is clear that, for all impact parameters, 
the timings $t_\mathrm{c}$ and $t_\mathrm{dyn}$ are all located 
near $t = 2.0$ Gyrs,
and have no clear dependence on $b$.

\begin{figure}[tb]
\includegraphics[width=\columnwidth]{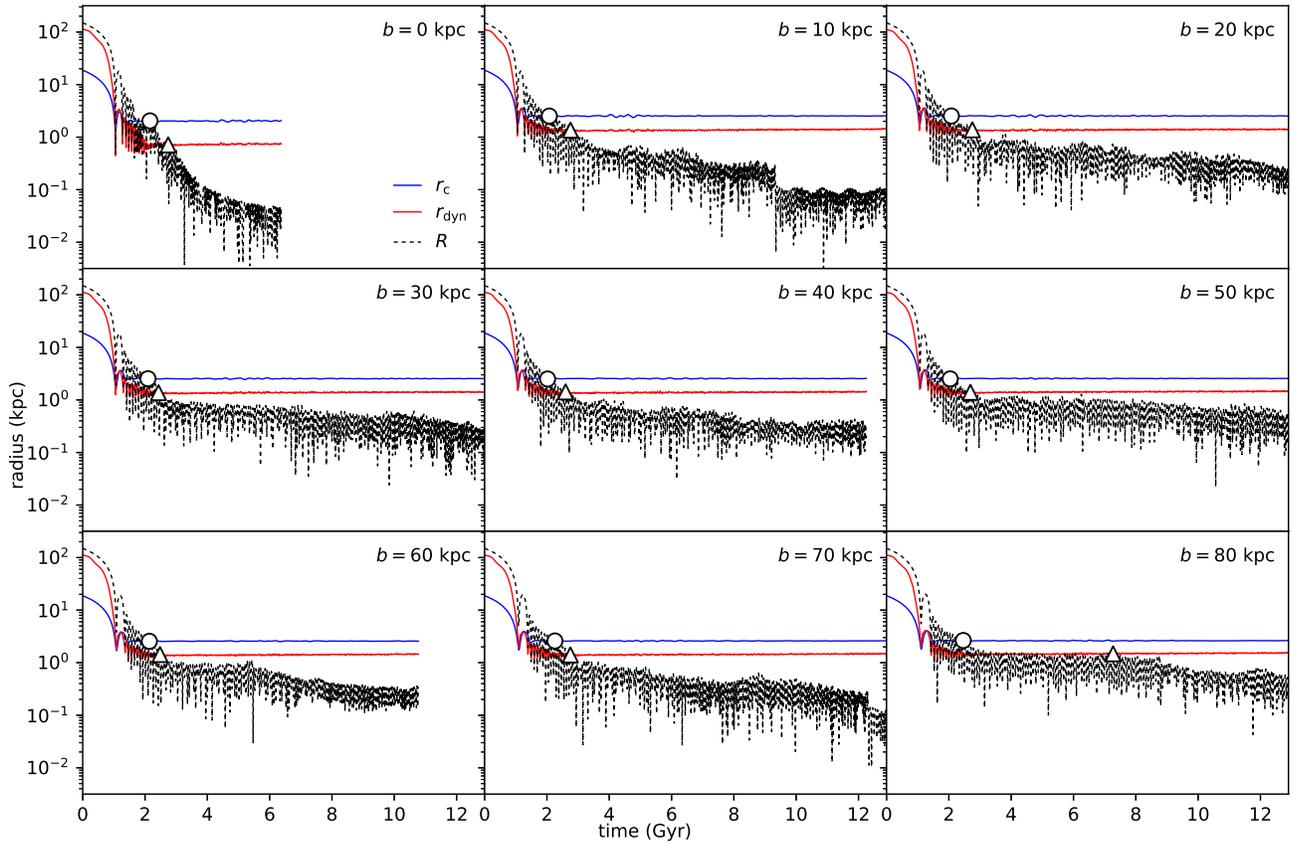}
\caption{Same as Fig.~\ref{R_c5} but for Model B. 
         The simulations are run until either a bound binary has been formed (See Fig.~\ref{energy}) 
         or an upper limit of 13 Gyr is reached.}
\label{R_c8}
\end{figure}

Fig.~\ref{R_c8} shows the similar analysis for Model B.
The time evolution of $r_\mathrm{c}$ and $r_\mathrm{dyn}$ are 
basically the same in both Model A and Model B,
except that the final values in Model B are 
slightly larger than those in Model A. 
Table~\ref{radii_tab} summarizes the final values of $r_\mathrm{c}$ and $r_\mathrm{dyn}$ for both models. 
Because the mass of the MBHB is the same ($M_\mathrm{BB} = 1.6 \times 10^{9} M_\odot$), 
a greater final value of $r_\mathrm{dyn}$ in Model B means a lower mean density within $r_\mathrm{dyn}$. 
In addition, we found that the final values of $r_\mathrm{c}$ in Model B are 
systematically greater than those of Model A by $\sim$ 0.5 kpc.
Since the total mass in both models are the same and $r_\mathrm{c}$ is 
derived from $r_\mathrm{e}$(see Equation(\ref{r_c_def})),
the larger values of $r_\mathrm{c}$ in Model B imply that 
the values of $r_\mathrm{e}$ in Model B are systematically larger 
and the core stellar density slope of the remnants in Model B are shallower than those in Model A. 

The evolution of $R$, however, shows a completely 
different feature in Model A and B. 
Comparing Fig.~\ref{R_c5} and Fig.~\ref{R_c8}, we notice that 
the orbital radii $R$ decrease much slower in Model B than 
those in Model A, 
and the dual MBHs remain for a longer period between 
$t_\mathrm{c}$ and $t_\mathrm{dyn}$ in Model B. 
Except for the mergers with the impact parameter $b = 0$ and $b = 10$, 
all other dual MBHs in Model B seem to slow down their orbital decay rate 
at a separation of $\sim 1$ kpc after $t_\mathrm{dyn}$. 
This is probably because the stellar particles originally surrounding the MBHs 
are scattered away during the merging process, 
therefore the efficiency of the dynamical friction decreases. 
The case of $b = 80$ even has $t_\mathrm{dyn} >$ 6\,Gyr. 

\begin{table*}
\small
\caption{Final values of the defined radii $r_\mathrm{c}$ and $r_\mathrm{dyn}$\label{radii_tab}}
  \begin{tabular}{@{}|c|c|c|c|c|c|c|c|c|c|c|@{}}
  \tableline
  $b$ & $0$ & $10$ & $20$ & $30$ & $40$ & $50$ & $60$ & $70$ & $80$ \\
  \hline
  $r_\mathrm{c, A}$ & 2.0293 & 2.0275 & 2.0034 & 2.0339 & 2.0396 & 2.0166 & 2.0630 & 2.0486 & 2.0914 \\
  \hline
  $r_\mathrm{c, B}$ & 2.0425 & 2.5199 & 2.5285 & 2.5324 & 2.5048 & 2.5150 & 2.5536 & 2.5915 & 2.6448 \\
  \hline
  $r_\mathrm{dyn, A}$ & 0.5799 & 0.6220 & 0.4693 & 0.5668 & 0.5394 & 0.5520 & 0.6243 & 0.6087 & 0.5088 \\
  \hline
  $r_\mathrm{dyn, B}$ & 0.71499 & 1.3629 & 1.3815 & 1.3760 & 1.4238 & 1.3992 & 1.4312 & 1.4096 & 1.4510 \\
  \tableline
  \end{tabular}
  \tablenotetext{a}{Table of the final core radii and dynamical radii of the merger remnant. 
 $r_\mathrm{c, A}$ and $r_\mathrm{c, B}$ are the core radii of Model A and Model B, respectively.
  Similarly, $r_\mathrm{dyn, A}$ and $r_\mathrm{dyn, B}$ are the dynamical radii of the two models.}
 \tablenotetext{b}{All units are in kpc.}
  \label{radii_tab}
\end{table*}

After the MBHs reach and remain within $r_\mathrm{dyn}$, 
the dual MBHs continued to dissipate their orbital energy 
to the surrounding particles via dynamical friction, to
eventually form a bounded MBHB. 
However, during the evolution, 
the dual MBHs may oscillate between bound and unbound state 
due to the interaction with the stellar and dark halo particles. 

The MBHB's orbital energy $E_\mathrm{BB}$ is defined as
\begin{equation}
    E_\mathrm{BB}(t) = \frac{2(K + U)}{M_\bullet},
    \label{t_bound_def}
\end{equation}
where $K$ is the MBHB's kinetic energy, and $U$ is the potential energy. 
Considering the MBHB as an isolated system, 
we derive $K$ by removing the center of mass velocity of the MBHB, 
and neglect the potential of other particles. 
The final timing when $E_\mathrm{BB} = 0$ is defined 
as the bound time, $t_\mathrm{b}$. 

\begin{figure}[tb]
\includegraphics[width=\columnwidth]{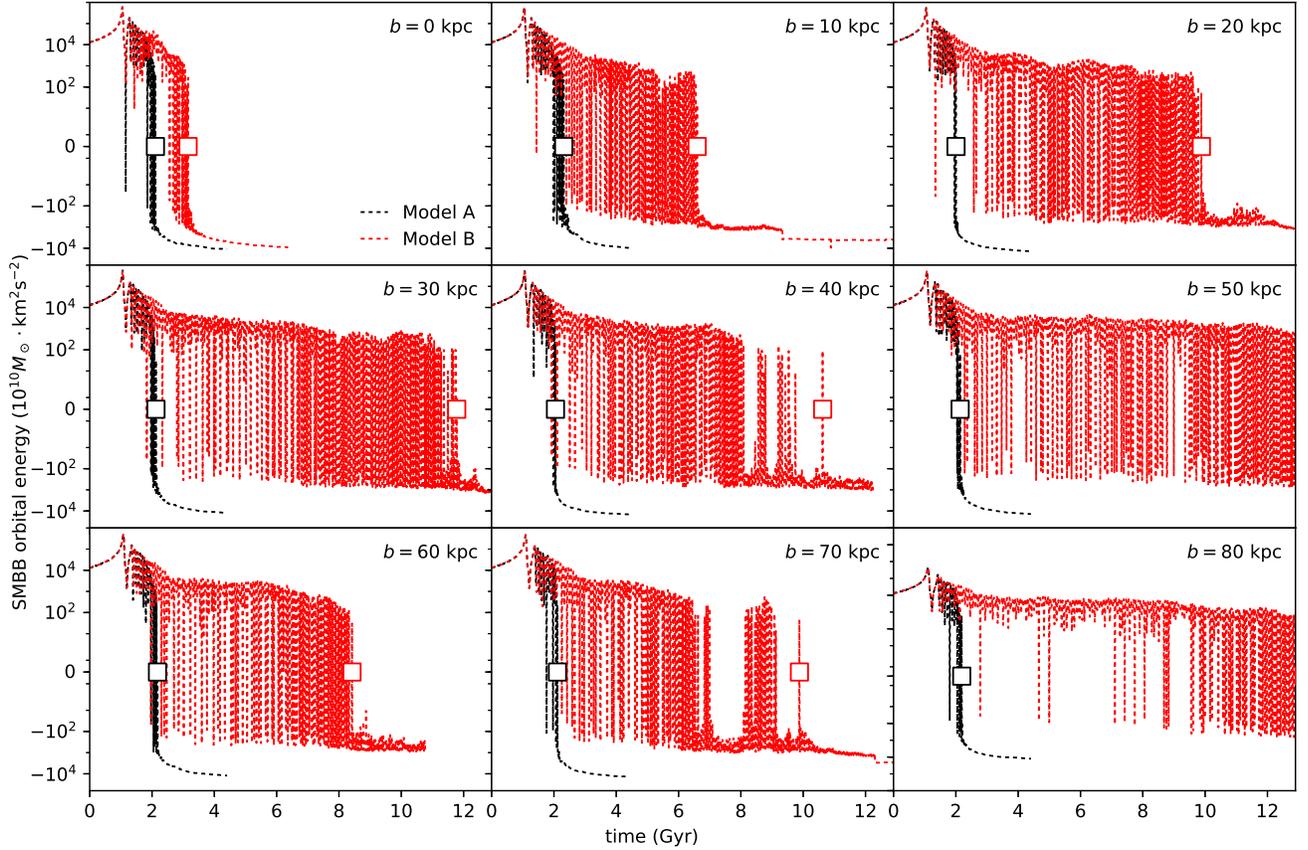}
\caption{Time evolution of the MBHB orbital energy $E_\mathrm{BB}$. 
         The impact parameter $b$ is shown in 
         the upper-right corner of each panel.
         The black dotted line represents $E_\mathrm{BB}$ for Model A,
	and the red dotted line is for Model B.
	The black and red squares indicate $t_\mathrm{b}$, 
	i.e. the final timings when $E_\mathrm{BB} = 0$, 
        in Model A and Model B, respectively.
	Note that the mergers with the impact parameter 
        $b = 50$ and $b = 80$ could not form a bound binary 
	within 13 Gyr.}
\label{energy}
\end{figure}

Fig.~\ref{energy} shows the time evolution of the MBHB orbital energy
for both Model A and Model B. 
The black and red squares indicate the final timings 
when $E_\mathrm{BB}$ drops to zero, 
i.e. the bound time $t_\mathrm{b}$, in Model A and Model B, respectively. 
In Model A, the orbital energy goes up and down and finally becomes negative.
Clearly, $t_\mathrm{b}$ is right after $t = 2$ Gyr
for all mergers with nine different impact parameters.
This result, together with Fig.~\ref{R_c5}, shows 
that this timing $t_\mathrm{b}$ always follows 
after $t_\mathrm{dyn}$, 
in agreement with the previous 
studies\citep{2006RPPh...69.2513M, 2018MNRAS.477.2310B}.
However, for Model B,
not all dual MBHs could form a bound binary within $13$\,Gyr. 
Only the head-on merger shows similar evolution as in Model A, but
the other cases have dual MBHs alternating 
between bound and unbound for very long periods. 

The fact of having a large time difference between 
$t_\mathrm{b}$ and $t_\mathrm{dyn}$ in Model B 
is contrary to the previous studies that show $t_\mathrm{b} \approx t_\mathrm{dyn}$
\citep{2002MNRAS.331..935Y, 2006RPPh...69.2513M}, 
and has significant implications.  
First, it implies the fundamental difference between the definitions of $E_\mathrm{BB}$ and $r_\mathrm{dyn}$.  
$E_\mathrm{BB}$ describes a dynamical state of the dual MBH system, 
and does not depend on the surrounding environment directly. 
However $r_\mathrm{dyn}$ is defined through the 
enclosed stellar mass compared with the MBHB mass only, 
which has nothing to do with the dynamics of the dual MBHs. 
Secondly, our simulations focus on different combinations of 
density structure of the progenitors and merger geometries. 
Thus, the discrepancy between $t_\mathrm{b}$ and $t_\mathrm{dyn}$ 
reflects the fact that the evolution of the MBHB is 
a complicated process affected by many factors. 
In Model A, a denser stellar core surrounding the initial MBH enables the core 
to maintain its original structure for a longer period 
during the merging process. 
Therefore a stronger dynamical friction could drive 
the dual MBHs to a closer distance, 
and eventually form a bound MBHB at the radius $\sim r_\mathrm{dyn}$.
While in Model B, a larger stellar core with a shallower central 
stellar density is adopted. 
Since the merger remnant in Model B has a more diffuse central stellar density, 
a longer time is required for dynamical frictions to diminish the energy 
of the dual MBHs until forming a bound pair. 
Even after the MBHs reach the radius 
$r_\mathrm{dyn}$ of $1.36 \sim 1.45$ kpc, their orbital radius 
continue reducing 
but the two MBHs remain unbound.
 
\begin{figure}[tb]
\includegraphics[width=\columnwidth]{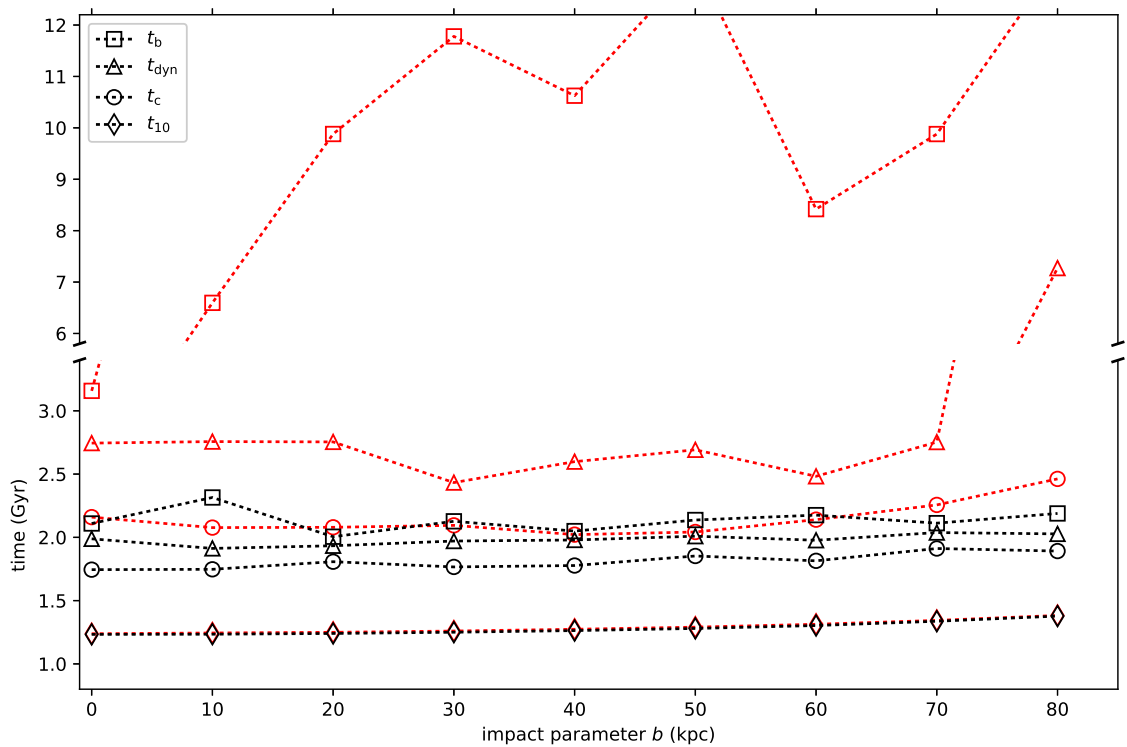}
\caption{The important timings over different impact parameters. 
  	Please see the main text for the definitions of $t_\mathrm{10}$, 
	$t_\mathrm{c}$, $t_{\mathrm{dyn}}$, and $t_\mathrm{b}$.
	The black markers are for Model A, and the red markers 
	are for Model B. Note that the $b = 50$ and the $b = 80$ cases
	both have $t_\mathrm{b} > 13$ Gyr, therefore are not shown 
	in the figure.}
\label{time_over_b}
\end{figure}

The above results show that the effects of the
stellar density structure of the progenitor and the merger geometry 
on the evolutionary timescale of the MBHB formation are actually a 
coupled situation, in a way that both factors should be 
considered simultaneously.
To provide complete information about the MBH sinking and 
the MBHB formation in our simulations,  
all timings defined in this subsection are summarized in 
Fig.~\ref{time_over_b}. 
Furthermore, in order to have more idea about the early stage of MBH sinking,
the timing when the MBH orbital radius equals 10 kpc, i.e. $t_{10}$, 
is also determined and plotted in Fig.~\ref{time_over_b}.

In general, for Model A,
the timings have no significant dependence on
impact parameters. 
Only the merger with $b = 10$ forms an MBHB slightly later 
than the other cases
and has the relatively longer duration 
between $t_\mathrm{dyn}$ and $t_\mathrm{b}$, 
up to a few hundreds of Myr.
To investigate this, we have further tested several more simulations 
with $b$ near 10 kpc, from $b = 6$ to $b = 18$ kpc. 
The results of the characteristic timings are shown 
in Fig.~\ref{time_over_b_sup}. 
It can be seen that the deviation of $t_\mathrm{b}$ from $t_\mathrm{dyn}$ is
clear for most mergers with impact parameters 
between $b = 8$ kpc and $b = 18$ kpc.
It is also shown that the deviation exhibits significant variations.

As for Model B, although all the $t_{10}$ coincides with those of Model A,
the $t_\mathrm{c}$ differs from that of Model A significantly.
There is a slight trend for $t_\mathrm{c}$ to increase with $b$
when $b > 40$ kpc, but the most prominent signature of Model B is a significantly longer dynamical friction timescale 
for the dual MBHs to evolve from $t_\mathrm{dyn}$ to $t_\mathrm{b}$.

\subsection{The MBHB Orbital Properties} \label{orbit} 

To further investigate the MBHB orbital properties,
the MBHB angular momentum is discussed here.
Although the MBHB forms after $t_\mathrm{b}$, 
the angular momentum is still calculated and presented from $t=0$
as it could give hints about the MBH orbits during the MBHB formation process. 

\begin{figure}[tb]
\includegraphics[width=\columnwidth]{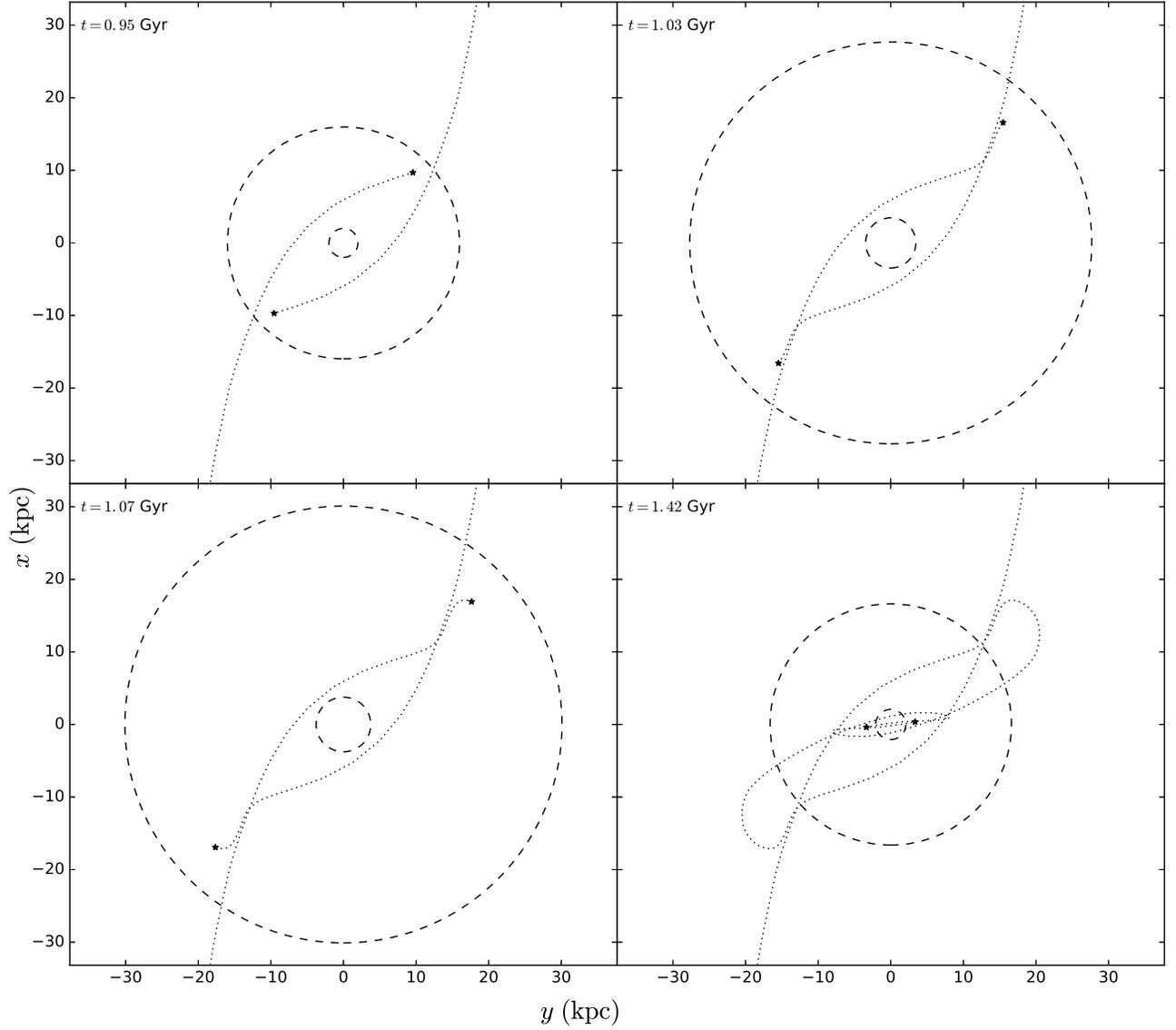}
\caption{The trajectories of two MBHs on the $xy$-plane 
         for the mergers with the impact parameter $b = 50$ kpc. 
         The time of each snapshot is shown on the top-left corner. 
         Two black asterisks indicate the positions of two MBHs, 
           and the dotted curves show their trajectories. 
           The larger dashed circle indicates  
           the effective radius $r_\mathrm{e}$, and the smaller dashed circle  
           shows the core radius $r_\mathrm{c}$.}
\label{b50}
\end{figure}

To give an example of MBH orbits,
Fig.~\ref{b50} presents the trajectories of two MBHs for 
the case of $b = 50$\,kpc in Model A.
From $t = 0.95$ Gyr (top-left) to $t = 1.07$ Gyr (bottom-left), 
it can be seen that
the MBHB orbit encounters a turn-over, 
such that the moving directions of the MBHs change drastically. 
The orbital direction of the MBHs first switch 
from clockwise to counter-clockwise, 
and then back to clockwise. 

\begin{figure}[tb]
\includegraphics[width=\columnwidth]{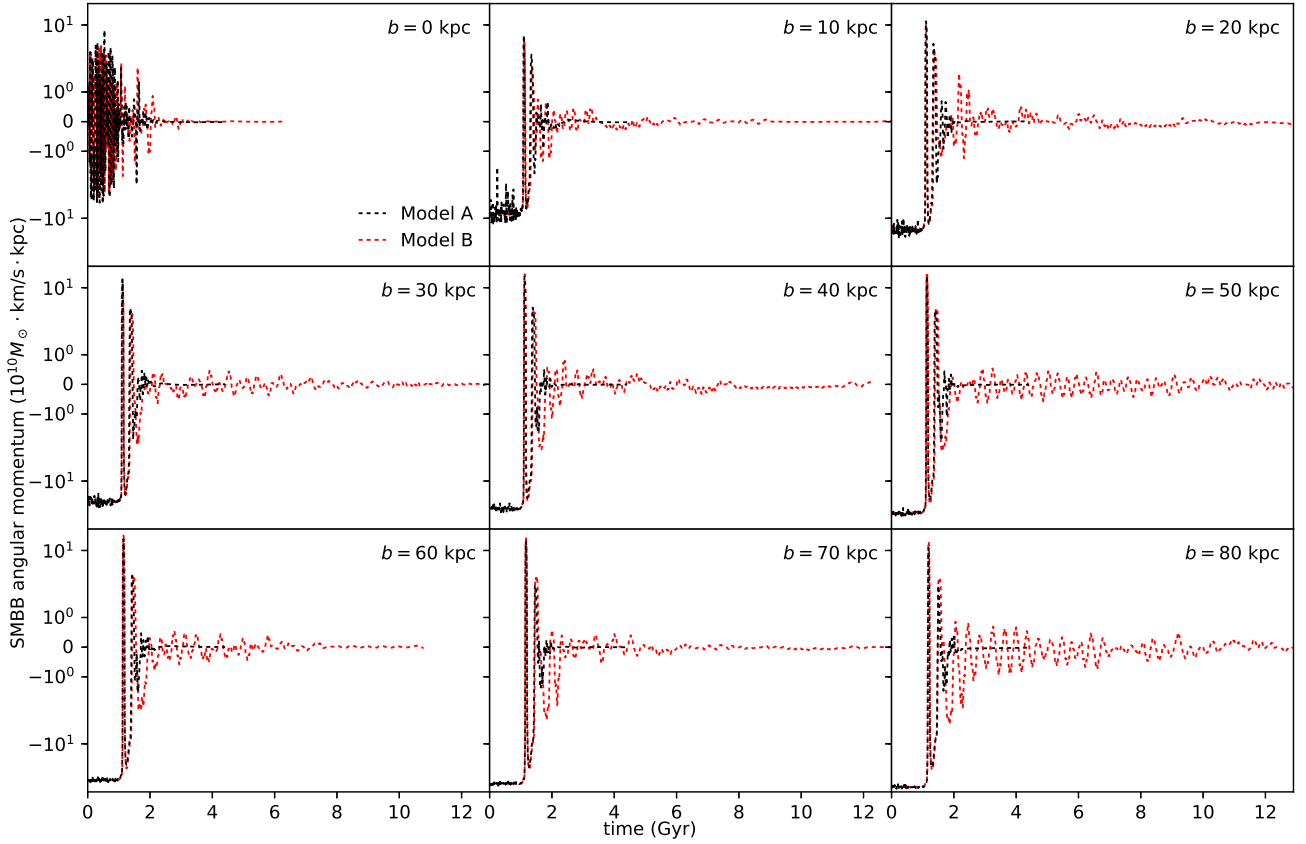}
\caption{Time evolution of the MBHB angular momentum $L_\mathrm{BB}$. 
	Since the initial MBHB orbital plane lies on the $xy$-plane, 
	$L_\mathrm{BB}$ is along the $z$-axis. 
	 The impact parameter $b$ is shown in 
         the upper-right corner of each panel.
         The values of Model A are shown in black, 
         and the values of Model B are shown in red.
         }
\label{angmom}
\end{figure}

The MBHB angular momentum is defined as
\begin{equation}
    \mathbf{L}_\mathrm{BB} \equiv M_\mathrm{BB} \sum_{i}^{1, 2} \mathbf{r}_i \times \mathbf{v}_{i},
    \label{LBB_def}
\end{equation}
where $\mathbf{r}_i, \mathbf{v}_i$ are the position  
and velocity vectors of two MBHs relative to their barycenter. 
Fig.~\ref{angmom} shows the time evolution of the MBHB angular momentum. 
The MBHB orbital plane  lies on the $xy$-plane, and hence 
only the $z$-component of the $\mathbf{L}_\mathrm{BB}$ 
is shown in the figure. 
Since the initial velocity as well as the initial separation of 
two MBHs are fixed in all mergers, 
the total energy of two MBHs are the same initially 
in all simulations. 
However, due to the difference in impact parameters, 
the initial angular momentum $\mathbf{L}_\mathrm{BB}$ are not the same. 
Fig.~\ref{angmom} shows that, for the case of $b = 0$,  
the angular momentum $L_\mathrm{BB}$ is zero at the beginning 
of the simulation and then
oscillates quickly between positive and negative values.
For the other cases of Model A, the angular momentum is 
negative initially.
After two major sign-switchings at $t = 1-2$ Gyr, 
$L_\mathrm{BB}$ decreases steadily and converges to zero.
Furthermore, together with Fig.~\ref{b50}, we can see that 
for the merger with $b = 50$ kpc,
the first major sign-switching of $L_\mathrm{BB}$
happens after the first encounter of two MBHs at $t = 0.95 - 1.03$ Gyr, 
as shown in the top two panels of Fig.~\ref{b50}. 
In addition, the second major sign-switching happens around $t = 1.43$ Gyr, 
as shown in the bottom right panel of Fig.~\ref{b50}. 
After that,
both MBHs sink into the center of the merged system.

As for Model B, the evolution of $L_\mathrm{BB}$
basically follows that of the trend in Model A, of the same $b$.
However, except for the merger with $b = 0$, all other cases in Model B
show significant oscillations after two major sign-switchings. 
It means that dynamical friction from the stellar background
is not efficient at removing angular momentum from 
$L_\mathrm{BB}$, and 
therefore the dual MBHs may stay unbound for a much longer period.
 
\subsection{The Triaxiality}\label{stellar} 

To investigate the effect of impact parameters on the
structures of merged cores, the triaxiality of the stellar core 
is determined by the method of the 
moment-of-inertia tensor \citep{1991ApJ...378..496D, 2006ApJ...642L..21B}. 
For the region within the core radius, an ellipsoidal density distribution 
is set as
\begin{equation}
    \rho \equiv \rho(\zeta),\ \rm{with}\ \zeta = \left( x_1^2 + \frac{x_2^2}{\eta^2} + \frac{x_3^2}{\xi^2}\right)^\frac{1}{2}, 
    \label{rho_ellipse}
\end{equation}
where $\zeta$ is the elliptical radius, 
and $\eta$ and $\xi$ are the axial ratios with $\xi \le \eta \le 1$. 
The $\eta$ and $\xi$ can be determined through the calculations of the moment-of-inertia tensor 
and then rotate the coordinates of the stellar particles in the core 
(within $r_\mathrm{c}$).
We compute the inertia tensor iteratively until both 
differences between the input 
and output of $\eta$ and $\xi$ 
are smaller than $10^{-5}$.
After $\eta$ and $\xi$ are determined, the triaxiality index 
$T$ is calculated as \citep{2006RPPh...69.2513M, 2018MNRAS.477.2310B}  
\begin{equation}
    T \equiv \frac{1 - \eta^2}{1 - \xi^2}. 
    \label{triaxiality_index}
\end{equation}
They are only evaluated after $t_\mathrm{c}$ when 
the MBHs are in the core of the merged system. 

\begin{figure}[tb]
\includegraphics[width=\columnwidth]{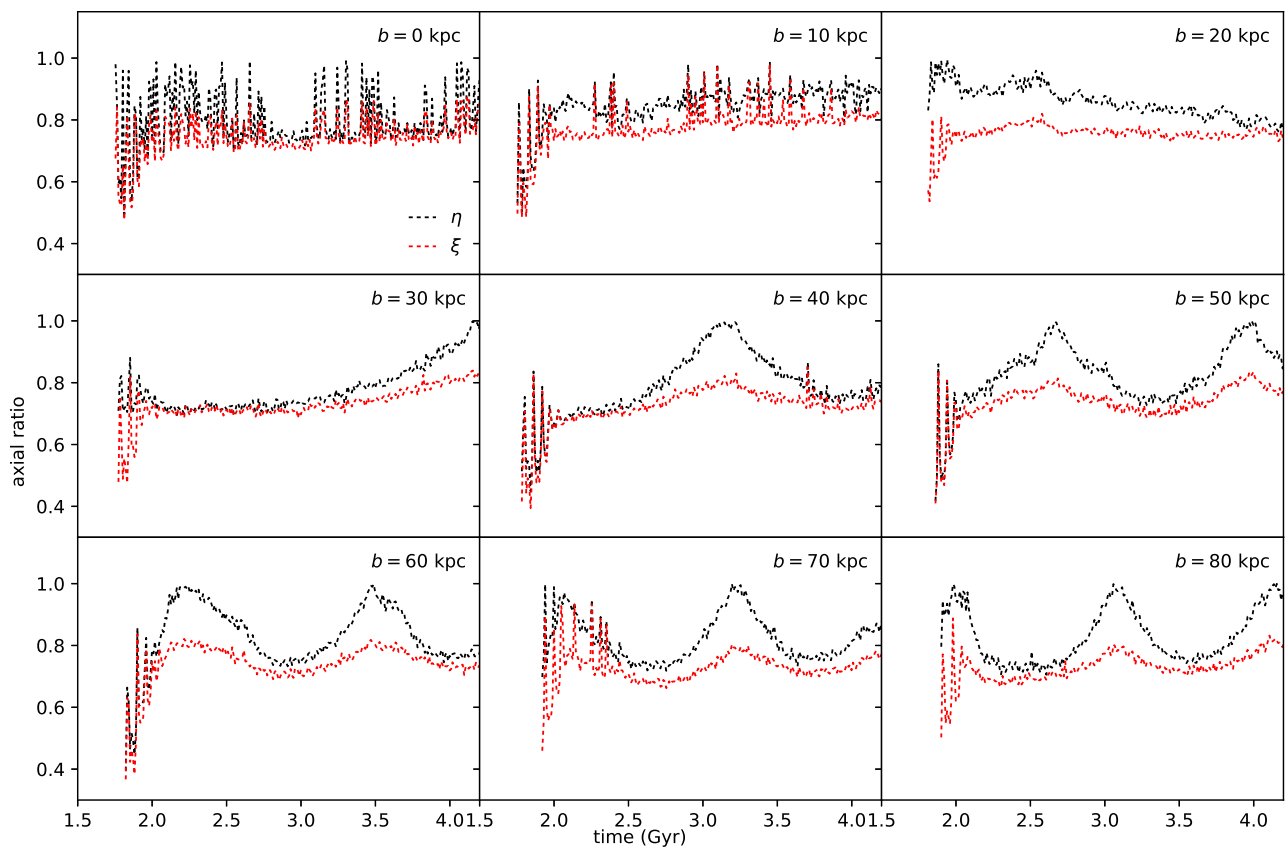}
\caption{Time evolution of the axial ratio $\eta$ and $\xi$
	for Model A.
         The impact parameter $b$ is shown in 
         the upper-right corner of each panel.
         The $\eta$ is shown in black, 
         and $\xi$ in red.
         }
\label{axialratio_c5}
\end{figure}

\begin{figure}[tb]
\includegraphics[width=\columnwidth]{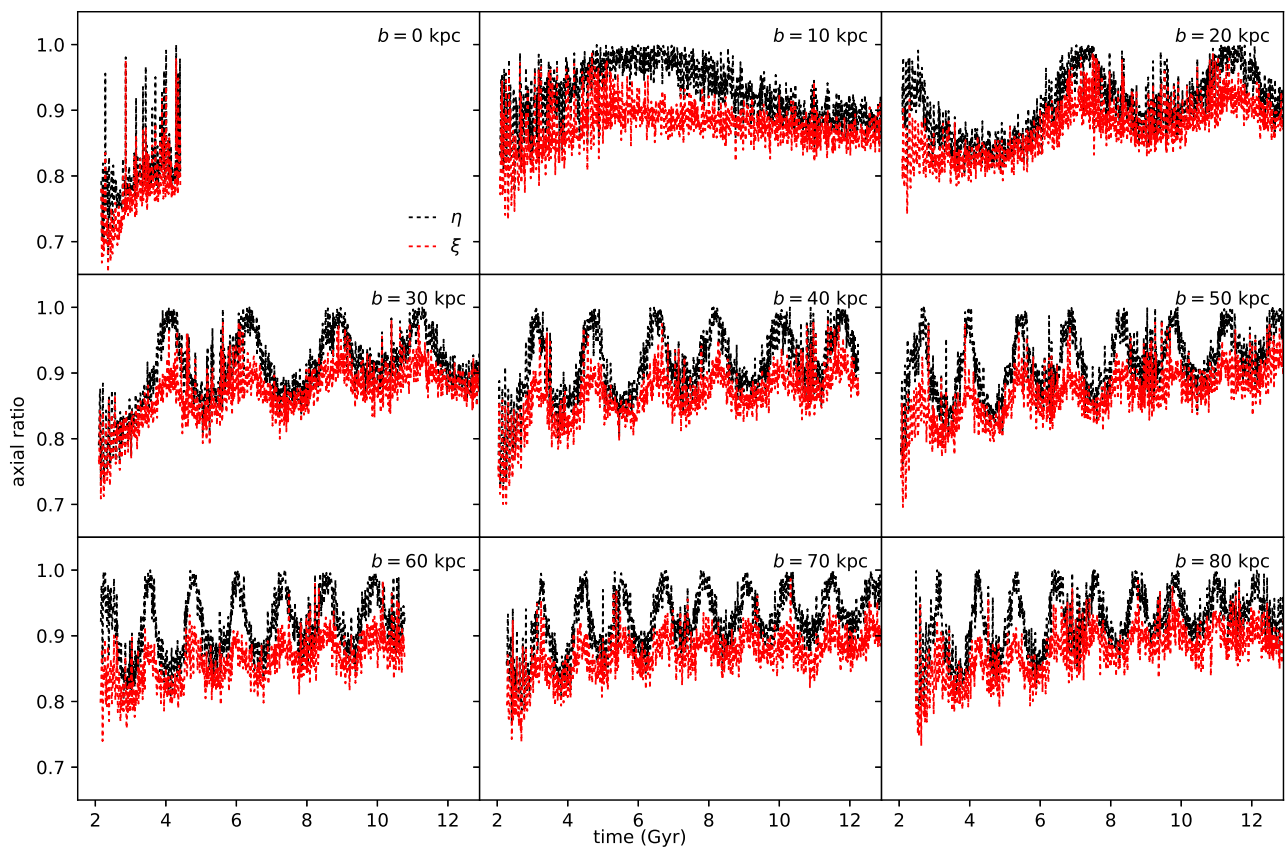}
\caption{Same as in Fig.~\ref{axialratio_c5} 
	but for Model B. 
         }
\label{axialratio_c8}
\end{figure}

Fig.~\ref{axialratio_c5} and Fig.~\ref{axialratio_c8} present 
the time evolution of the axial ratios 
$\eta$ and $\xi$ for different impact parameters of 
Model A and Model B, respectively.
The most prominent difference between Model A and Model B \textbf{is} that 
$\eta - \xi$ is significantly greater in Model A. 
This implies that the core structure is close to the shape of 
an oblate ellipsoid in Model A.
Since the progenitor in Model A has a more concentrated stellar center, 
the oblateness of the remnant core can be expected 
if the original stellar density structure surrounding the original MBH 
is better preserved during the merging process. 
On the other hand, the smaller $\eta - \xi$ in Model B suggests 
that the remnant stellar core is closer to a spheroid. 
Secondly, $\eta$ and $\xi$ are systematically 
closer to unity in Model B than in Model A. 
Note that $\eta = \xi = 1$ gives a spherical symmetric core, 
thus having $\eta$ and $\xi$ closer to unity suggests that
the remnant core structures in Model B are closer to spherical.

In terms of impact parameters, 
$b$ mainly governs the dynamical properties of the evolving core,
such that mergers with the same $b$ would 
result in their remnant cores having similar motion.
For the case of $b=0$,
both $\eta$ and $\xi$ oscillate violently all the time.
As for $b=10$, there still many peaks of oscillations are observed,
but the difference between $\eta$ and $\xi$ becomes clearer.
For all the remaining cases with larger impact parameters, 
the axial ratios vary gently.

\begin{figure}[tb]
\includegraphics[width=\columnwidth]{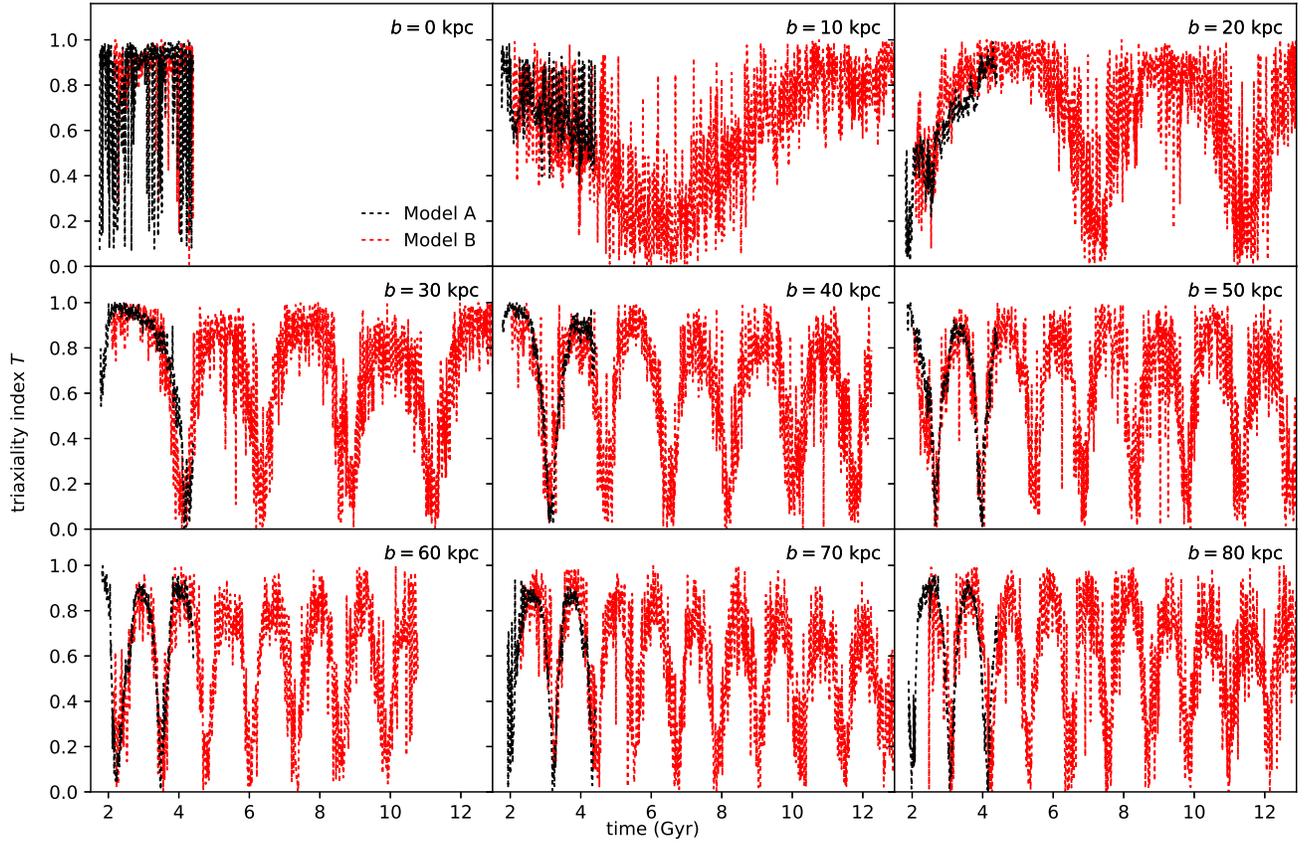}
\caption{Time evolution of the triaxiality index $T$. 
	The impact parameter $b$ is shown in 
        the upper-right corner of each panel.
        The black dotted curves are for Model A, 
        and the red dotted curves are for Model B.
         }
\label{triaxial}
\end{figure}

The time evolution of triaxiality index $T$ is shown in Fig.~\ref{triaxial}. 
The stellar core is considered to be oblate if $0 \le T < 0.5$ 
and prolate if $0.5 < T \le 1$, while $T = 0.5$ indicates 
the maximum triaxiality.
It can be seen that the evolution of $T$ follows the same trend for
Model A and Model B, i.e. the black lines 
and the red lines basically
overlap with each other in Fig.~\ref{triaxial}.
However, for different impact parameters, 
similar to the axial ratios in
Fig.~\ref{axialratio_c5} and Fig.~\ref{axialratio_c8}, 
the triaxiality indexes in the mergers with 
$b = 0$ and $b = 10$ oscillate violently.
In addition, the merger with $b = 10$ has its triaxiality index at 
$T \approx 0.5$ for the longest duration.

\begin{figure}[tb]
\includegraphics[width=\columnwidth]{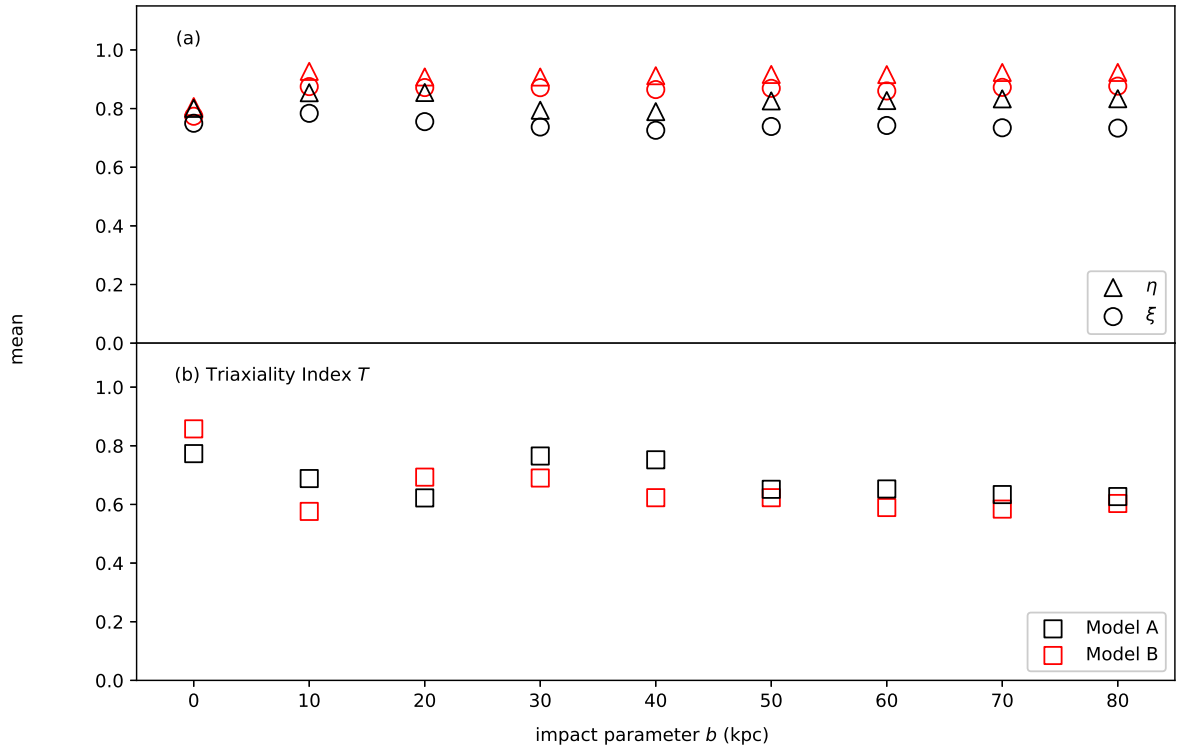}
\caption{(a)Time average of the axial ratios $\eta$ and $\xi$ 
             as a function of impact parameter $b$.
             The triangle and circle indicate $\eta$ and $\xi$, respectively. 
         (b)Time average of the triaxiality index $T$ 
             as a function of impact parameter $b$.
             For both panels, black markers are for Model A, 
             and red markers are for Model B.
          }
\label{mean_triaxial}
\end{figure}

To have an overview of the above evolutions,
Fig.~\ref{mean_triaxial} shows
the mean values of $\eta$, $\xi$, and $T$ 
as a function of the impact parameter.
The black markers are for Model A, and the red markers are for Model B.
Panel (a) shows that remnants in Model B generally have a higher 
value of axial ratios $\eta$ and $\xi$ when compared with Model A, 
except for $b = 0$. 
However, such dependencies are not found on the mean values of $T$.

In general, the periodic variations of $\eta$, $\xi$, and $T$ 
are caused by the rotation of the remnant core.
Since the evolution of $T$ in Model A and Model B clearly overlaps, 
this variation in $T$ only depends on the impact parameter.

\section{Discussions}\label{sec:discussions}

Based on the results presented in Section~\ref{sec:results}, 
we discuss the relations between different impact parameters 
and the half-mass radii of the progenitor galaxies here.
Note that the half-mass radius $r_{1/2}$ of the progenitor galaxies 
in Model A and Model B are $12.07$ kpc and $19.31$ kpc, respectively. 
Because all progenitors have the same mass, 
a greater value of $r_{1/2}$ means a lower 
average stellar density within $r_{1/2}$. 

\subsection{Head-on Mergers} \label{subsecs:head-on} 

During a head-on merger ($b = 0$), 
two MBHs move toward each other and enter the density peak 
of the merged system directly. 
They approach each other nearly along a straight line and thus 
the initial angular momentum $L_\mathrm{BB}$ is close to zero. 
However, due to the strong interaction between the MBHs and other particles, 
as shown in Fig.~\ref{angmom}, 
the angular momentum $L_\mathrm{BB}$ 
keeps alternating between positive and negative values, 
which would be impossible if both MBHs were 
only governed by their two-body gravitational force. 
This strong interaction between the MBHs and other particles also makes the 
triaxiality index $T$ varying rapidly all the time, 
as can be seen from Fig.~\ref{triaxial}. 
Nevertheless, because the initial angular momentum $L_\mathrm{BB}$ is zero, 
with enough dynamical friction from other particles, 
MBHs sink toward the center of the merging system 
smoothly as presented in Fig.~\ref{R_c5} and Fig.~\ref{R_c8}. 
Furthermore, the evolution of MBHs during the head-on merger shows 
little dependence on 
the models of the progenitor galaxies. 
The evolution of the MBHs in Model A and Model B are basically 
the same although the timings $t_\mathrm{c}$, $t_\mathrm{dyn}$ and $t_\mathrm{b}$ 
in Model B are slightly longer than in Model A. 
This is reasonable since the progenitors in Model B have a lower 
stellar density at their central region and therefore 
the dynamical friction is not as efficient as in Model A.

\subsection{Mergers with Non-zero Impact Parameters}\label{subsecs:b>0} 

When $b > 0$, 
because the initial angular momenta of the progenitors are not zero, 
the merged remnant core becomes more triaxial, 
and the discrepancies between $\eta$ and $\xi$ 
are generally larger than those in a head-on merger.

Furthermore, in Model A, we notice that 
the difference between $t_\mathrm{b}$ and $t_\mathrm{dyn}$ in the merger with 
$b = 10$ kpc is larger than that in the mergers with larger $b$, 
as shown in Fig.~\ref{time_over_b}.
We believe the cause of this is the combination of a small half-mass radius 
of the progenitor galaxies and a small impact parameter $b$. 
In Model A, since the progenitors have a smaller half-mass radius, 
their stellar particles are more concentrated around the central MBHs. 
When the impact parameters are small, 
the two progenitors will have their half-mass radii overlapping with each other and therefore 
result in strong interaction between the stellar particles during the merging process. 
Meanwhile, each MBH moves directly into the high-density region 
of the other progenitor, and suffers strong perturbations 
from the stellar particles. 
Hence, $t_\mathrm{b}$ and the difference 
between $t_\mathrm{b}$ and $t_\mathrm{dyn}$ increase, 
as seen in the merger with $b=10$ kpc. 
However, if the impact parameters are significantly larger 
than the half-mass radius of the progenitors, 
the stellar particles will remain near each MBH for a longer period, 
and continue to subtract the orbital energy of the MBHs. 
We will discuss the variation of $t_\mathrm{b}$ in Model A 
in more details in Section~\ref{subsecs:b=10}.

On the other hand, as shown in Fig.~\ref{time_over_b}, 
the timing $t_\mathrm{b}$ is much later than $t_\mathrm{dyn}$ 
in Model B, comparing with in Model A.
In the cases $b = 50$ and $b = 80$ kpc in Model B, 
their $t_\mathrm{b}$ are even longer than our simulation time. 
In Model B, the progenitors have a larger half-mass radius 
and a lower density stellar core. 
Although the half-mass radii of the progenitors also overlap, 
dynamical frictions are less efficient in subtracting the 
orbital energy of the MBHs because 
the two progenitors merge into a core with lower stellar density. 
Meanwhile, since there are less stellar particles near each MBH, 
the original stellar particles surrounding the MBH get scattered 
away more easily, leaving the two MBHs "naked". 
The orbital radii of the MBHs will then continue 
to shrink by interacting with the 
background stars at a much lower rate, 
as shown in Fig.~\ref{R_c8}, and therefore 
the timings $t_\mathrm{b}$ becomes much later than $t_\mathrm{dyn}$.
In addition, the timings $t_\mathrm{b}$ and $t_\mathrm{dyn}$ in Model B 
show similar variations as in Fig.~\ref{time_over_b_sup}. 
The MBHs' orbits are 
perturbed by the direct collision between the progenitor cores, 
and less stellar particles are around the MBHs, 
it is thus even harder for the MBHs to form a bound binary.

It can also be seen from Fig.~\ref{mean_triaxial} 
that mergers in Model B leave 
rounder stellar cores in the orbital plane when compared with those in Model A, 
such that $\eta$ and $\xi$ are systematically closer to unity.

In general, by comparing Fig.~\ref{R_c5} with Fig.~\ref{R_c8}, 
it is clear that the orbital evolutions between Model A and 
Model B are quite different for non-zero impact parameters.
In Model A, the orbital radii are able to shrink smoothly 
after $t_\mathrm{dyn}$, 
and reaching saturation at values of a few times $0.01$ kpc. 
While in Model B, the respective final values are 
generally close to $0.1$ kpc. 
This suggests that the dynamical friction timescale could be much longer 
in mergers of progenitors with low central density, and that MBHs could stay at 
hundreds-of-parsec separations much longer in these mergers.
However, simulations with a larger number of stellar particles are required to 
resolve the three-body interactions between the stellar particles and the MBHs, 
which are beyond the scope of this work.

\subsection{Model A with Impact Parameters Near 10 kpc}\label{subsecs:b=10} 

\begin{figure}[tb]
\includegraphics[width=\columnwidth]{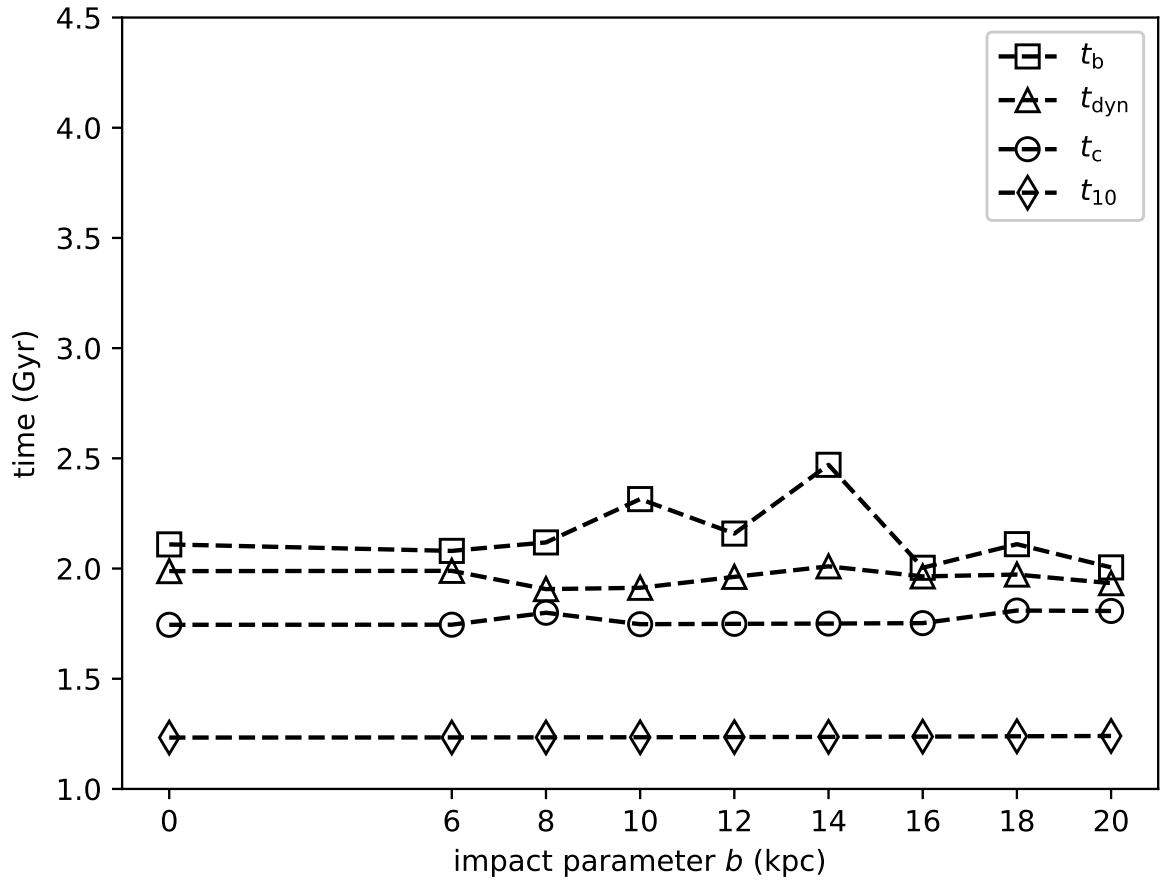}
\caption{Same as in Fig.~\ref{time_over_b},
	but for mergers with $b$ ranging from 6 to 18 kpc. 
	The simulations are supplementary to Model A 
	in order to investigate the peculiar timescale shown 
	at the $b = 10$ case.}
\label{time_over_b_sup}
\end{figure}

Lastly, it is worth noticing that in Model A, 
$t_\mathrm{b}$ fluctuated
near $b \approx 10$ kpc as shown in Fig.~\ref{time_over_b_sup}. 
A possible cause is that, when the two progenitors 
approach each other under small impact parameters, 
the MBH goes into the high-density part of another galaxy directly, 
which leads to strong interactions between the MBHs 
and the surrounding stellar particles, 
causing more perturbed MBH orbits than the other cases. 
Note that the perturbations might be numerical, since there 
is no general trend in the distribution of $t_\mathrm{b}$ for $b$ 
between 10 kpc and 20 kpc. 
To estimate the scale of the perturbations, 
we run the $b = 10$ kpc case in Model A 
for three times using the same identical setup, i.e. we build progenitors using 
the same density profile and performed merger simulations with $b = 10$ kpc. 
We obtained the standard deviations for 
$t_\mathrm{c}$, $t_\mathrm{dyn}$, and $t_\mathrm{b}$ 
to be $1.57$, $7.49$, and $3.02$ Myrs, respectively. 
The scale of perturbations are much smaller than the variations 
observed in the distribution 
of $t_\mathrm{b}$ in Fig.~\ref{time_over_b_sup}.
Future refined simulations, such as increasing the number of particles 
or using more accurate force integration techniques, are required 
to confirm the MBHB orbital evolutions in these cases.
 
\section{Conclusions}\label{sec:conclusion} 

N-body simulations with more than 1.6 million particles 
are employed to investigate 
the MBHB formation during a merger of two galaxies. 
With different impact parameters and different half-mass 
radii of the progenitor galaxies, 
we study the orbital evolution of the MBHs 
from an initial separation of 300 kpc 
until the formation of a bound MBHB.
We categorize the evolution of the MBHs into different stages according to 
the important timings $t_\mathrm{c}$, $t_\mathrm{dyn}$ and $t_\mathrm{b}$, 
and also present the time evolution of the MBH orbital radius, 
the MBHB orbital angular momentum, 
and the triaxiality of the merged stellar core. 

Our results show that, for mergers of progenitor galaxies with a denser stellar core, 
the timings $t_\mathrm{c}$, $t_\mathrm{dyn}$ and $t_\mathrm{b}$ are earlier than 
those of progenitors with a less-denser stellar core, which is expected because dynamical friction is stronger 
when the MBHs are moving within a denser stellar environment, 
therefore the orbital radius of the two MBHs shrinks faster. 
In addition, we see that the timing $t_\mathrm{dyn}$ could deviate from $t_\mathrm{b}$ greatly, 
especially when the progenitors have less-denser stellar core. 
The deviation between $t_\mathrm{dyn}$ and $t_\mathrm{b}$ are large 
in mergers of progenitors with less-denser stellar core. 
Furthermore, the final values of the MBHB orbital radii are also different, 
such that mergers with less-denser stellar cores generally have greater final values than 
those with denser ones by an order.

Another important finding in this work is that, 
when the impact parameters are small in Model A, 
the MBHs would collide directly within the core radius of 
the progenitor galaxies, 
and the dual MBHs would suffer strong perturbations from the stellar particles, 
which leads to variations in the timings $t_\mathrm{b}$ for impact parameters 
between 10 kpc and 20 kpc. This suggests that resolving the later stages of the 
MBHB under higher stellar density in the progenitors is more difficult 
if the MBHs would collide directly into the central region with nonzero initial 
angular momentum.

There
are other physical processes at play that could be important. 
Our work provides a bigger picture of the foundations of the 
later stages of the MBHB
orbital evolutions. Future works using a larger number 
of particles and more refined 
integration accuracy will be important in resolving 
the later stages of the MBHB evolution. 
Moreover, 
further relevant processes such as multiple mergers, 
gas accretion, 
heating or scouring effect by the MBHB formation 
should also be considered.

\section*{Acknowledgment}
We are thankful to the referee for very helpful suggestions.
This work is supported in part by 
the Ministry of Science and Technology, Taiwan,
under Ing-Guey Jiang's 
Grants MOST 105-2119-M-007-029-MY3, and MOST 106-2112-M-007-006-MY3.


\clearpage

\end{document}